\documentclass[runningheads]{llncs}
\usepackage[T1]{fontenc}
\usepackage{graphicx}
\usepackage{array}
\usepackage[numbers,sectionbib]{natbib}
\usepackage{hyperref}
\usepackage{multirow}
\usepackage{booktabs}
\usepackage{soul}
\usepackage{color}
\newcolumntype{P}[1]{>{\raggedright\arraybackslash}p{#1}}
\usepackage{bbding}
\begin{document}
\title{Flexible metadata harvesting for ecology using large language models}
\author{Zehao Lu\inst{1,3} \and
Thijs L van der Plas\inst{1,3} \Envelope \and
Parinaz Rashidi\inst{2} \and
W Daniel Kissling\inst{2} \and
Ioannis N Athanasiadis\inst{1}}
\authorrunning{Lu*, Van der Plas* et al.}
\institute{Wageningen University \& Research, 6708 PB Wageningen, The Netherlands\\
\email{\{zehao.lu,thijs.vanderplas,ioannis.athanasiadis\}@wur.nl}\and 
University of Amsterdam, 1090 GE Amsterdam, The Netherlands \and
These authors contributed equally.}

\maketitle              % typeset the header of the contribution
\begin{abstract}
Large, open datasets can accelerate ecological research, particularly by enabling researchers to develop new insights by reusing datasets from multiple sources. However, to find the most suitable datasets to combine and integrate, researchers must navigate diverse ecological and environmental data provider platforms with varying metadata availability and standards. To overcome this obstacle, we have developed a large language model (LLM)-based metadata harvester that flexibly extracts metadata from any dataset's landing page, and converts these to a user-defined, unified format using existing metadata standards. We validate that our tool is able to extract both structured and unstructured metadata with equal accuracy, aided by our LLM post-processing protocol. Furthermore, we utilise LLMs to identify links between datasets, both by calculating embedding similarity and by unifying the formats of extracted metadata to enable rule-based processing. Our tool, which flexibly links the metadata of different datasets, can therefore be used for ontology creation or graph-based queries, for example, to find relevant ecological and environmental datasets in a virtual research environment.

\end{abstract}

\section{Introduction}
Ecological and environmental sciences are vital to addressing the global biodiversity crisis, for example, by monitoring species diversity and ecosystem health, and by quantifying the effects of climate change, human interventions and conservation action \citep{johnston2025north, keck2025global, langhammer2024positive}. 
Increasingly, the scientific community has advocated that these challenges can only be met by integrating different data streams, including Earth observation, citizen science records, sensor networks, long-term ecological monitoring surveys and environmental DNA \citep{culina2018navigating, farley2018situating, hampton2013big, kissling2018towards, nathan2022big, vanderplas2025monitoring, runting2020opportunities}. 
Essential to this mission are best practices in the large-scale collection, sharing and re-analysis of environmental and ecological data \citep{farley2018situating, hampton2013big, kissling2018towards}. 
\begin{figure}[t]
    \centering
    \includegraphics[width=\linewidth]{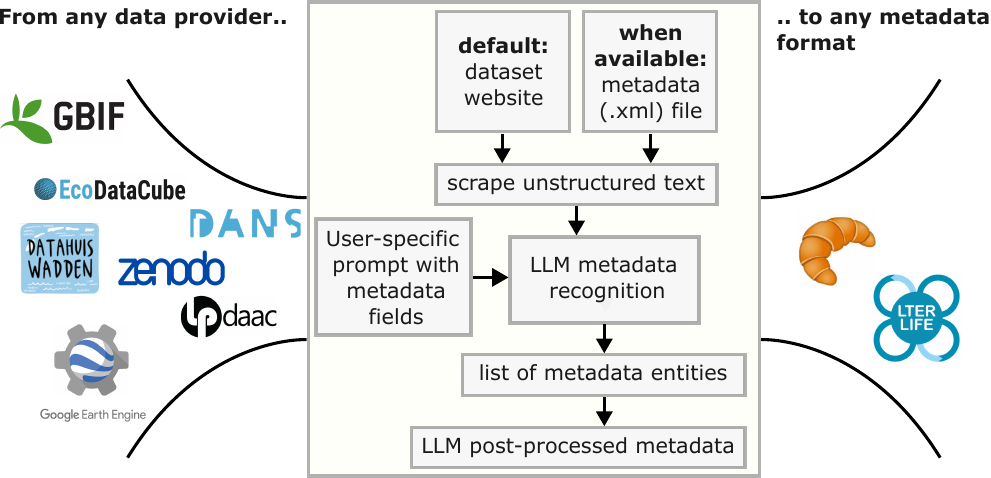}
    \caption{Diagram of our LLM-based metadata harvester that retrieves and converts metadata from any data provider to any metadata format.}
    \label{fig:diagram}
\end{figure}

For example, FAIR (findable, accessible, interoperable and reproducible) data facilitate efficient re-analysis \citep{wilkinson2016fair}, stored either by individual researchers at data archival providers such as GBIF (Global Biodiversity Information Facility) and Zenodo, or at organisational repositories such as Google Earth Engine. 

Metadata that describe the key data properties are critical to (using) FAIR data \citep{kissling2018towards, loffler2021dataset, wilkinson2016fair}. 
With metadata, researchers can determine the relevance of a dataset for their research purposes prior to analysing it, enabling them to efficiently search across a large set of datasets. However, while data providers often enable users to access or search across metadata from their catalogue, there is no unified way of retrieving and querying metadata across different data providers, hindered by differences in metadata availability and formatting (but note that catalogues of multiple data repositories exist \citep{culina2018navigating, noy2019google}). 
For example, interpreting metadata remains a key challenge for reusing FAIR data \citep{papoutsoglou2023benefits}. 
Such obstacles are especially challenging in ecology, where a wide range of data can be of relevance, distributed across different data repositories, involving several scientific disciplines \citep{culina2018navigating, loffler2021dataset}. While, in contrast, searching for scientific articles across different journals has become an anchor point of modern scientific research, searching for scientific datasets remains a challenge for most researchers \citep{Gregory2020Lost}. 
 
Metadata standards (or schemas) that unify metadata formats of different datasets (and data providers) are therefore crucial. However, different metadata standards are not always intercompatible due to differences in which metadata fields are used and how they are formatted, even if they are considered FAIR \citep{loffler2021dataset}. Importantly, there is no one-size-fits-all solution: data providers and researchers from different communities may need to prioritise different metadata (\textit{i.e.} different fields or different standards), just as academic journals vary in citation style and lay-out. Instead, a flexible solution is required that can retrieve metadata from any data provider and convert these to any user-defined metadata format. 

Large language models (LLMs) have been suggested to improve the accessibility of ecological datasets \citep{rafiq2025generative}. 
Indeed, LLMs have been used successfully to annotate textual, descriptive metadata such as title, description, keywords and domain \citep{zhang2023utilising, hayashi2024metadata, watanabe2025capabilities}, and to convert text input into ontologies \citep{caufield2024structured}. Yet, to maximise usability, an integrated, open-source LLM tool is needed that harvests metadata directly from dataset source web pages, including exact (numeric) metadata such as spatial resolution or license information, and returns these in a unified format.

Here, we introduce an LLM-based metadata harvester that scrapes metadata from any data provider and converts these to a user-specified format (Fig \ref{fig:diagram}). We evaluate our tool on an annotated set of datasets across seven providers and two output formats, and show that our LLM metadata harvester successfully retrieves both structured and unstructured metadata. Finally, we demonstrate how these retrieved metadata can be used to construct a metadata knowledge base that establishes links between datasets across providers. With this, our tool enables researchers to create user-specific knowledge bases of datasets that are designed for efficient data discovery, thereby facilitating `big data' ecology \citep{farley2018situating, runting2020opportunities}.

\section{Methods}
\paragraph{Named entity recognition}
We have developed an LLM-based approach that, when given an URL, automatically scrapes all metadata from any machine-readable text format, and converts these to any metadata format using an LLM that performs named entity recognition (Fig \ref{fig:diagram}). First, all text is scraped from a dataset landing page, and if available, from structured metadata files (such as \texttt{.xml} files). Next, the user specifies what metadata \textit{entity types} (\textit{i.e.}, fields that specify metadata types such as \textit{Title}, \textit{Publication date}, etc.) should be retrieved, including their definitions. This is integrated into a prompt (adapted from \citep{guo2025lightragsimplefastretrievalaugmented}) that asks the LLM to retrieve all \textit{entities} of these entity types. 
Because LLMs sometimes deviate from predefined schemas when generating structured outputs \cite{wang2025slotstructuringoutputlarge}, we use a second LLM call to post-process the extracted metadata entities and improve the formatting of the final output. Here, we task the LLM to constrain the output such that for each dataset, only 1 entity per entity type is retrieved. While this is generally applicable for most metadata entities (\textit{e.g.}, there can only be a single `publication date'), this forces other entity types to return an enumeration of (sub-)entities in a single text string (\textit{e.g.}, multiple authors are returned as as single entity). 
% \usepackage{multirow}
\begin{table}[h!]
\caption{Metadata fields considered for harvesting, using two different formats (LTER-LIFE and Croissant) with partial overlap. Notably, even where metadata fields overlap, their definitions (\textit{e.g.} standards and vocabularies) still differ: Croissant uses Croissant definitions, while LTER-LIFE definitions used here originate from DCAT-AP or ISO 19115, see table.}
\label{tab:metadatafield_overview}
\begin{tabular}{clll}
\toprule
\multicolumn{1}{l}{\textbf{Group}}        & \textbf{Metadata field}  & \textbf{Croissant?} & \textbf{LTER-LIFE?} \\
\midrule
\multirow{3}{*}{Metadata on metadata}     & Metadata date            & -                   & Yes - ISO 19115                    \\
                                          & Metadata language        & Yes - Croissant                & Yes - ISO 19115                    \\
                                          & Responsible organization & -                   & Yes - ISO 19115                    \\
\midrule 
\multirow{5}{*}{Identification}           & Title                    & Yes - Croissant                 & Yes - DCAT-AP                      \\
                                          & Description              & Yes - Croissant                 & Yes - DCAT-AP                      \\
                                          & Unique Identifier        & -                   & Yes - DCAT-AP                      \\
                                          & Resource type            & -                   & Yes - DCAT-AP                      \\
                                          & Keywords                 & Yes - Croissant                 & Yes - DCAT-AP                      \\
\midrule 
\multirow{3}{*}{Data contact information} & Data creator             & Yes - Croissant                 & Yes - DCAT-AP                      \\
                                          & Data contact point       & -                   & Yes - DCAT-AP                      \\
                                          & Data publisher           & Yes - Croissant                 & Yes - DCAT-AP                      \\
\midrule 
\multirow{3}{*}{Spatial properties}       & Spatial coverage         & -                   & Yes - DCAT-AP                      \\
                                          & Spatial resolution       & -                   & Yes - DCAT-AP                      \\
                                          & Spatial reference system & -                   & Yes - ISO 19115                    \\
\midrule 
\multirow{2}{*}{Temporal properties}      & Temporal coverage        & -                   & Yes - DCAT-AP                      \\
                                          & Temporal resolution      & -                   & Yes - DCAT-AP                      \\
\midrule 
\multirow{2}{*}{Intellectual rights}      & License                  & Yes - Croissant                 & Yes - DCAT-AP                      \\
                                          & Access rights            & -                   & Yes - DCAT-AP                      \\
\midrule 
\multirow{6}{*}{Distribution}             & Distribution access URL  & -                   & Yes - DCAT-AP                      \\
                                          & Distribution format      & -                   & Yes - DCAT-AP                      \\
                                          & Distribution byte size   & -                   & Yes - DCAT-AP                      \\
                                          & Same as                  & Yes - Croissant                 & -                            \\
                                          & Date published           & Yes - Croissant                 & -                            \\
                                          & Date last modified       & Yes - Croissant                 & -                           \\
\bottomrule
\end{tabular}
\end{table}

\paragraph{Metadata field descriptions}
Users specify which metadata the LLM should extract, by inserting the metadata fields (entity types) and definitions (entity descriptions, \textit{e.g.}, using existing metadata standards and vocabularies) in the prompt. To demonstrate this, we considered two different metadata standards: the LTER-LIFE\footnote{LTER-LIFE is a research infrastructure project to develop Digital Twins of ecosystems (\url{https://lter-life.nl/en}). As part of this, a metadata standard (currently version 0.0.1) has been developed, which we will refer to here as `LTER-LIFE'.} metadata format and Croissant, which are used by the ecology and machine learning research communities, respectively. 
Firstly, we used the LTER-LIFE metadata format, consisting of a community-developed minimum set of 21 metadata fields across 7 categories (see Table \ref{tab:metadatafield_overview}). LTER-LIFE metadata definitions make use of the Dutch ISO19115-based geography metadata profile\footnote{\url{https://geonovum.github.io/Metadata-ISO19115/}} and the DCAT3 (Data Catalog Vocabulary version 3) standard vocabulary\footnote{\url{https://www.w3.org/TR/2024/REC-vocab-dcat-3-20240822/}}. We translated Dutch definitions to English and manually supplemented entity descriptions where the standard definitions were insufficient. 
Secondly, Croissant is a metadata format designed to make datasets `machine learning ready' \citep{akhtar2024croissant}. We used Croissant metadata definitions as listed under the `Required' and `Recommended' fields from `Dataset-level Information'\footnote{\url{https://docs.mlcommons.org/croissant/docs/croissant-spec.html}}. Croissant fields partially overlap with the LTER-LIFE metadata format (both require a title, license, etc.), but their metadata field definitions still differ. 

\paragraph{Software implementation}
We used Python packages \texttt{BeautifulSoup} and \texttt{Playwright} to scrape website landing pages, and \texttt{BeautifulSoup} to parse \texttt{.xml} files. LLM calls and subsequent analysis and visualisation were performed in Python 3. We used and compared the OpenAI GPT-4 \citep{openai2023gpt4} and Google Gemini 2.5 Flash-05-20 \citep{gemini2-5flash} LLMs. We adapted our prompt from LightRAG \citep{guo2025lightragsimplefastretrievalaugmented}, with the critical difference that we specify which entity types should be extracted (as defined by the desired metadata format), whereas the default LightRAG implementation extracts entities of any type. We calculated description embeddings using SBERT all-MiniLM-L6-v2 \citep{reimers-2019-sentence-bert}, which creates same-length embeddings regardless of the input text length, allowing us to compute the cosine similarity between pairs of embeddings. All code is available at \url{https://github.com/LTER-LIFE/llm-metadata-harvester}.

\begin{table}[b!]
\caption{Overview of datasets used in the evaluation, separated by seven (meta)data providers.}
\label{tab:dataset_overview}
\begin{tabular}{P{4.5cm}P{3cm} P{1.5cm} P{3cm}}  % Adjust widths as needed
\toprule
\textbf{Name} & \textbf{Provider} & \textbf{URL} & \textbf{Abbreviation} \\
\midrule
Dutch forest reserves database and network & DANS & \href{https://lifesciences.datastations.nl/dataset.xhtml?persistentId=doi:10.17026/dans-2bd-kskz}{page} & Dutch forest database \\
\midrule
Ecotopenkaart 2016 & Datahuis Wadden & \href{https://datahuiswadden.openearth.nl/geonetwork/srv/api/records/A0h06_NlSEuNlium5OO3FA}{page}, \href{https://datahuiswadden.openearth.nl/geonetwork/srv/api/records/A0h06_NlSEuNlium5OO3FA/formatters/xml}{xml} & Ecotope map 2016 \\
Ecotopenkaart 2017 & Datahuis Wadden & \href{https://datahuiswadden.openearth.nl/geonetwork/srv/eng/catalog.search#/metadata/L-mHomzGRuKAHGMkUPjY9g}{page} & Ecotope map 2017 \\
Waddenbalans 2024 & Datahuis Wadden & \href{https://datahuiswadden.openearth.nl/geonetwork/srv/eng/catalog.search#/metadata/0fe7e64b-50b3-4cee-b64a-02659fc2b6c7}{page} & Wadden balance 2024 \\
\midrule
Actual probability distribution for Quercus robur & EcoDataCube & \href{https://stac.ecodatacube.eu/veg_quercus.robur_anv.eml/collection.json?.language=en}{page} & Oak distribution \\
Cloud-free reconstructed Landsat bimonthly NDVI & EcoDataCube & \href{https://stac.ecodatacube.eu/ndvi_glad.landsat.ard2.seasconv/collection.json?.language=en}{page} & Landsat NDVI \\
Cloud-free reconstructed Landsat yearly blue band & EcoDataCube & \href{https://stac.ecodatacube.eu/blue_glad.landsat.ard2.seasconv.m.yearly/collection.json}{page} & Landsat blue \\
Cloud-free reconstructed Landsat yearly green band & EcoDataCube & \href{https://stac.ecodatacube.eu/green_glad.landsat.ard2.seasconv.m.yearly/collection.json}{page} & Landsat green \\
\midrule
Waterleidingduinen camera trap P1 & GBIF & \href{https://www.gbif.org/dataset/74196cd9-7ebc-4b20-bc27-3c2d22e31ed7}{page} & Camera trap P1 \\
Waterleidingduinen camera trap P2 & GBIF & \href{https://www.gbif.org/dataset/f9ba3c2e-0636-4f66-a4b5-b8c138046e9e}{page} & Camera trap P2 \\
Waterleidingduinen camera trap P3 & GBIF & \href{https://www.gbif.org/dataset/bc0acb9a-131f-4085-93ae-a46e08564ac5}{page} & Camera trap P3 \\
eBird observation dataset & GBIF & \href{https://www.gbif.org/dataset/4fa7b334-ce0d-4e88-aaae-2e0c138d049e}{page} & eBird \\
\midrule
Harmonized Landsat Sentinel-2 & Google Earth Engine & \href{https://developers.google.com/earth-engine/datasets/catalog/NASA_HLS_HLSS30_v002}{page} & HLS \\
\midrule
MODIS Terra MOD09A1 Version 6.1 & LP DAAC & \href{https://lpdaac.usgs.gov/products/mod09a1v061/}{page} & MODIS \\
\midrule
Downscaled LUH2 land use scenarios for Belgium & Zenodo & \href{https://zenodo.org/records/8319440}{page} & LUH2 Belgium \\
Waterleidingduinen camera trap P1-3 & Zenodo & \href{https://zenodo.org/records/11440456}{page} & Camera trap P1-3 \\
\bottomrule
\end{tabular}
\end{table}

\begin{figure}[b!]
    \centering
    \includegraphics[width=\linewidth]{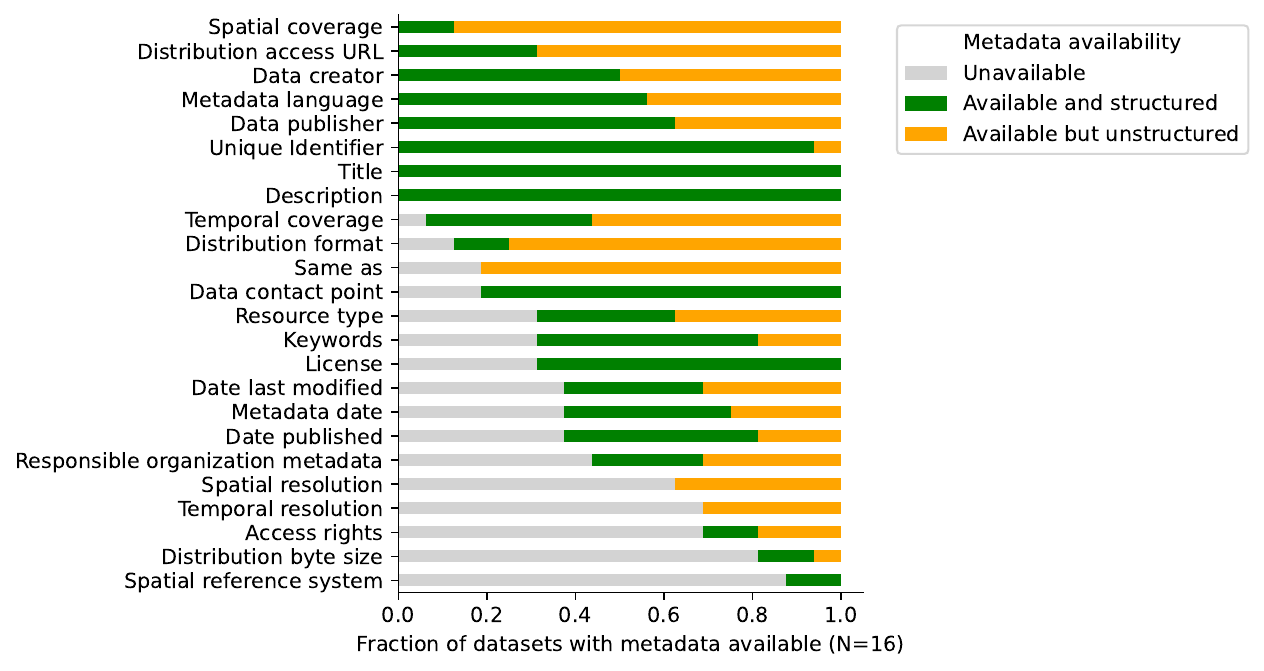}
    \caption{Overview of the manually annotated metadata for $N=16$ datasets (Table \ref{tab:metadatafield_overview}), which was either unavailable, available and structured, or available but unstructured.}
    \label{fig:metadata_availability}
\end{figure}

\paragraph{Evaluation datasets}
We evaluated our metadata harvester on 24 metadata fields, across 16 datasets from 7 data providers (see Tables \ref{tab:metadatafield_overview},\ref{tab:dataset_overview}). This set of datasets was selected to cover terrestrial and aquatic datasets, from both ecological and environmental domains, including \textit{in situ} observations (\textit{e.g.}, forest inventories, GBIF occurrence records) and Earth observation datasets (\textit{e.g.}, from MODIS, Landsat, Sentinel-2). One dataset (`Ecotopenkaart 2016', from Datahuis Wadden) also provided a metadata (\texttt{.xml}) file. Metadata varied in completeness across datasets, and was written in English for most datasets, except for the three datasets from Datahuis Wadden that were largely written in Dutch. (LLMs were prompted to return metadata fields in English.) 
We created a ground-truth dataset by manually annotating all metadata fields of both the LTER-LIFE and Croissant metadata formats for all 16 datasets. We annotated `N/A' when a metadata field was absent in the original source. In addition, we distinguished between unavailable metadata, present \textit{structured} metadata and present \textit{unstructured} metadata (Fig \ref{fig:metadata_availability}). We considered metadata to be structured if it was clearly labelled as metadata (\textit{e.g.}, a table with entry `Spatial resolution: 30 metres'), and to be unstructured if the information was only present in free-form text (\textit{e.g.}, if the spatial resolution would only be mentioned in a description that said `The data was aggregated yearly from 30-m bi-monthly Landsat data').

\paragraph{Evaluation metrics}
We distinguish between metadata fields that require a fuzzy match (\textit{Description} and \textit{Keywords}) and fields that require an exact semantic match (all other fields). Exact match fields enable us to use deterministic string-matching metrics, while fuzzy matches require LLM-based metrics. We use ROUGE-L F1 score, which compares string-to-string similarity, to test exact matches \citep{lin2004rouge}, and Faithfulness and Response Relevancy to test fuzzy matches. Faithfulness and Response Relevancy were originally designed for evaluating retrieval augmented generation (RAG) systems, which is conceptually similar to retrieving metadata from dataset landing pages. We assessed Faithfulness (fraction of claims in the metadata field that are supported by the dataset landing page) and Response Relevancy (similarity between generated questions that would yield the metadata as answer and the question ``What is the <FIELD> of this dataset?", where <FIELD> is `description' or `keywords') using Python package \texttt{Ragas} and the OpenAI GPT-3.5 Turbo LLM.

\paragraph{Statistics}
All scores are indexed between 0 (no match) and 1 (perfect match). Unless denoted otherwise, scores are presented with error bars as mean $\pm$ standard error of the mean (SEM). We only evaluate post-processed results unless specifically mentioned otherwise.
We use the likelihood-ratio test to test the effect of one variable while controlling for another. Specifically, we fit a mixed effects model using a fixed effect (variable) while controlling for a random effect (variable) and acquire the maximum log likelihood. We then obtain the maximum log likelihood of a second model using the random effect variable only, and test their log likelihood difference using the $\chi^2$-distribution to quantify the effect of the fixed effect variable \citep{mackenzie2018chapter3}.

\section{Results}
\subsection{Retrieving and converting metadata from any provider to any format}
\begin{figure}[h!]
    \centering
    \begin{minipage}[b]{0.6\linewidth}
        \includegraphics[width=\linewidth]{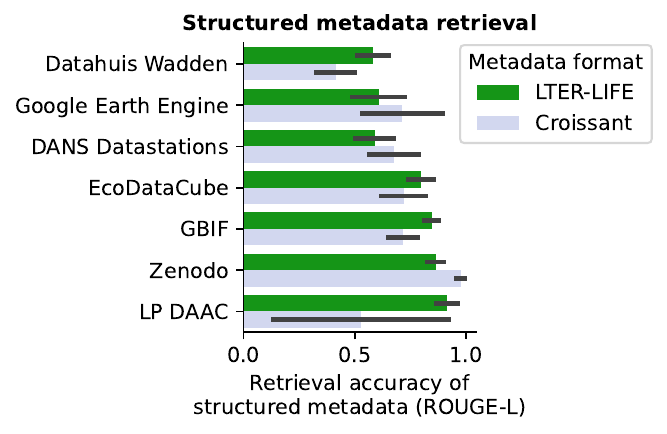}
    \end{minipage}%
    \hfill
    \begin{minipage}[b]{0.35\linewidth}
        \caption{
            Retrieval accuracy of post-processed, structured, present metadata per provider and metadata format (averaged across LLMs). Note that because the Croissant metadata format contains fewer fields, its uncertainty estimates are generally higher, especially for LP DAAC where N=2 for Croissant.
        }
        \label{fig:summary_rouge_per_provider}
    \end{minipage}
\end{figure}

Our metadata harvester successfully retrieved metadata from all 7 data providers and converted these into 2 different metadata formats (independently), with varying accuracy across data providers (Fig \ref{fig:summary_rouge_per_provider}). Note that the two different metadata formats (LTER-LIFE and Croissant) extract different metadata fields (Table \ref{tab:metadatafield_overview}), which largely explains the differences between their accuracy levels. Metadata retrieval accuracy varies significantly across data providers ($p=3 \cdot 10^{-6}$ for GPT-4, $p=9 \cdot 10^{-5}$ for Gemini 2.5 Flash, two-way likelihood-ratio test controlling for the random effect of metadata field), which is caused by differences across the datasets they host (Table \ref{tab:dataset_overview}).  

\begin{figure}[h!]
    \centering
    \includegraphics[width=\linewidth]{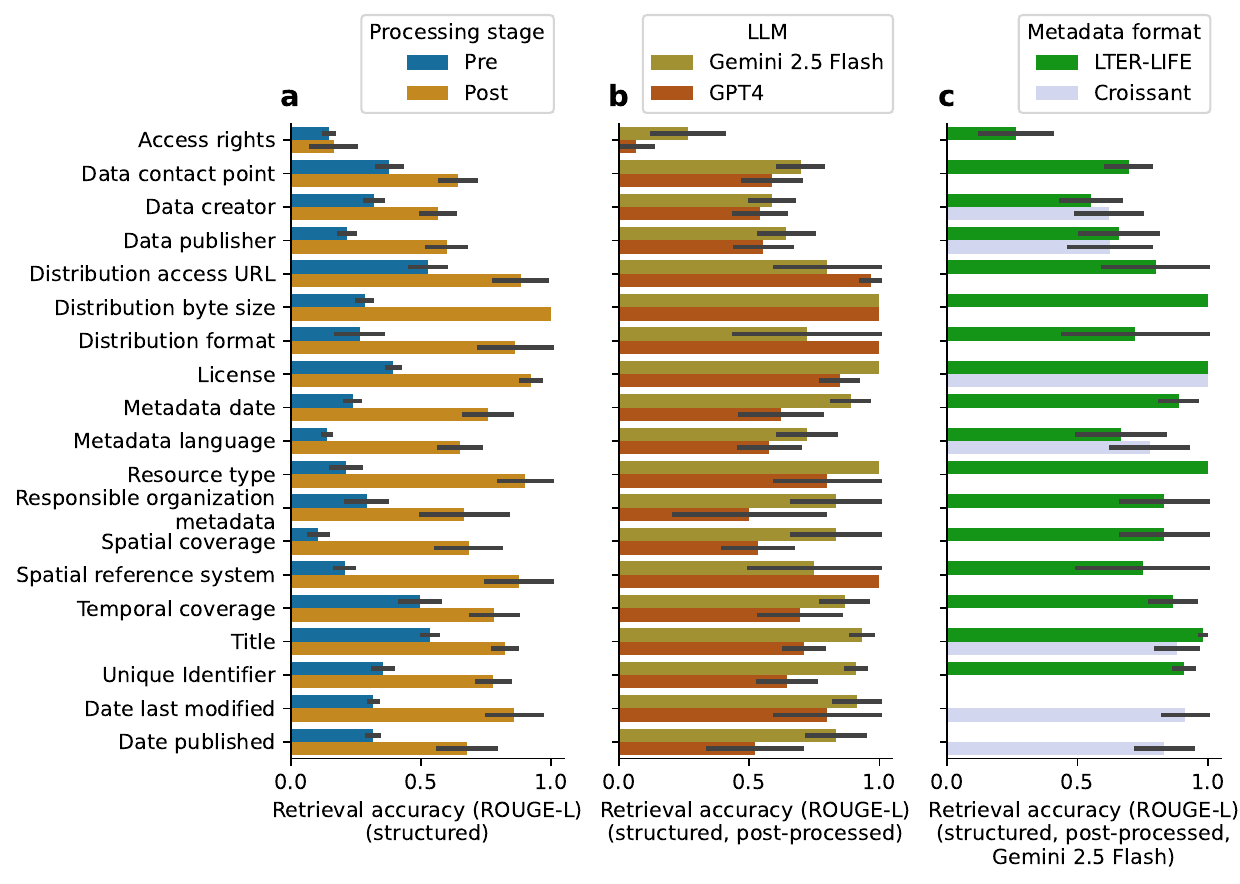}
    \caption{Retrieval accuracy of present, structured metadata, split by processing stage (a), LLM (b) and metadata format (c).}
    \label{fig:rouge_per_field}
\end{figure}

Our post-processing procedure - where a second LLM is prompted to adjust the extracted metadata if necessary to follow formatting guidelines - strongly improves accuracy across all metadata fields (Fig \ref{fig:rouge_per_field}a, $p=4 \cdot 10^{-62}$, two-way likelihood-ratio test controlling for the random effect of metadata field). LLM choice had a smaller effect, with Gemini 2.5 Flash outperforming GPT-4 (Fig \ref{fig:rouge_per_field}b, $p=7 \cdot 10^{-5}$, two-way likelihood-ratio test controlling for the random effect of metadata field). We observed that this difference occurred because GPT-4 occasionally failed to follow output instructions, changed metadata formatting, or failed to retrieve metadata.
The two metadata formats that we considered, LTER-LIFE and Croissant, extract partially overlapping sets of metadata fields (Table \ref{tab:metadatafield_overview}). Where these overlap, performance is similar (Fig \ref{fig:rouge_per_field}c, $p=0.91$, two-way likelihood-ratio test controlling for the random effect of all shared metadata fields), with (insignificant) variability caused by differences in metadata field definitions between the formats (that are used in the prompt).

\subsection{Filling in the blanks: recognising unstructured metadata}
\begin{figure}
    \centering
    \includegraphics[width=\linewidth]{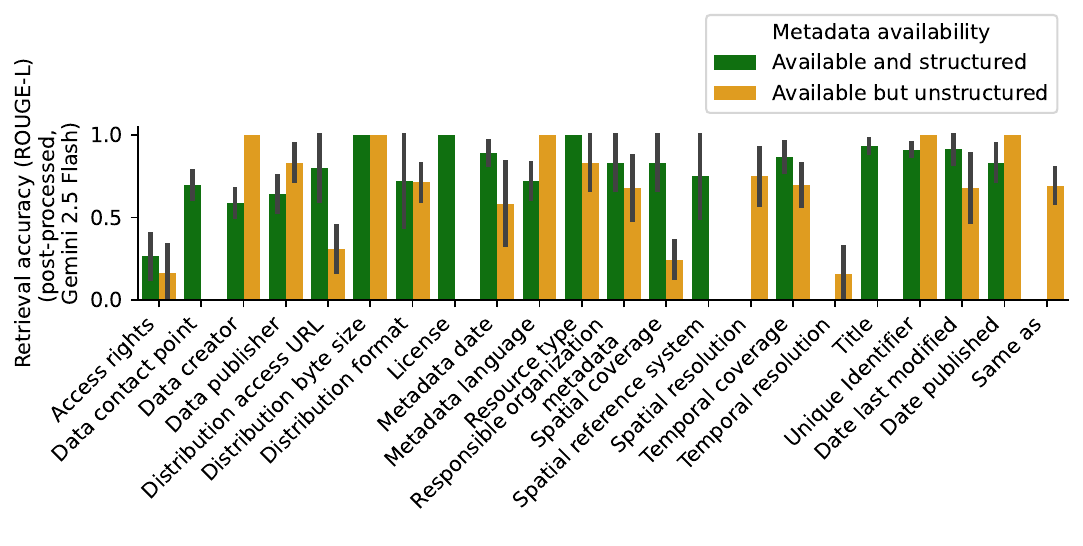}
    \caption{Retrieval accuracy of structured vs unstructured metadata using Gemini 2.5 Flash. Please note that for some metadata fields all metadata (across datasets) was either structured or unstructured (see Fig \ref{fig:metadata_availability}).}
    \label{fig:rouge_unstructured}
\end{figure}

The metadata harvester was able to recognise unstructured metadata with the same accuracy as structured metadata (Fig \ref{fig:rouge_unstructured}, $p=0.88$, two-way likelihood-ratio test controlling for the random effect of all shared metadata fields). Still, unstructured metadata scored lower on average for a number of metadata fields (Fig \ref{fig:rouge_unstructured}), in particular; \textit{Distribution access URL}, \textit{Metadata date} and \textit{Spatial coverage}. Unsurprisingly, these metadata fields are typically difficult to contextualise without structure: for example, while GBIF distribution access URLs are clearly structured by labelling them as "Endpoint: <URL>" in a metadata table (and as such, these are `structured' and were all retrieved perfectly), Datahuis Wadden URLs are listed without standardised labels, which could include the data endpoint, related data, reports, online data viewers etc. (as such, these are `unstructured' and scored 0.39 on average). 

\begin{figure}[bh]
    \centering
    \includegraphics[width=\linewidth]{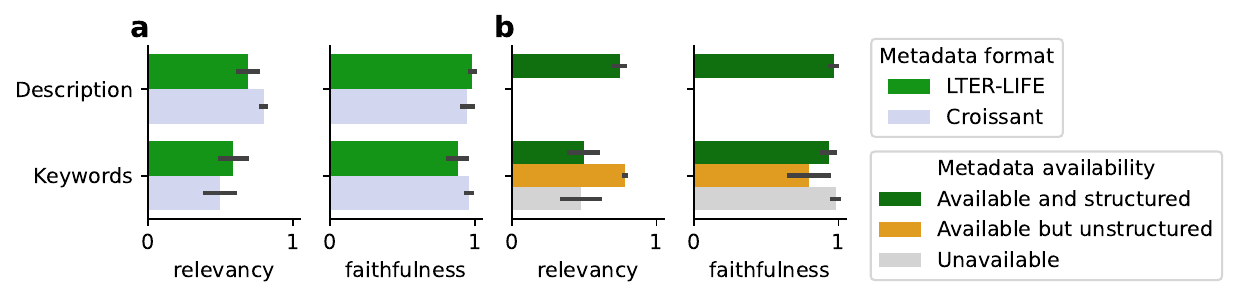}
    \caption{LLM-evaluated accuracy of \textit{Description} and \textit{Keywords} metadata. Only post-processed, Gemini 2.5 Flash results for \textit{Description} and \textit{Keywords} metadata was evaluated.}
    \label{fig:fuzzy_matches}
\end{figure}

We further evaluated the fuzzy match accuracy of the \textit{Description} and \textit{Keywords} fields, using LLM metrics Faithfulness and Relevancy (Fig \ref{fig:fuzzy_matches}). Keywords can be generated (especially if originally absent) and descriptions can be summarised, which prevent evaluation using the deterministic ROUGE metric. We found that generally Faithfulness was very high, meaning that the metadata was accurately describing the datasets. 

While retrieval accuracy does not differ significantly between structured and unstructured metadata, unstructured metadata is more often reported as `not available' than structured metadata (Table \ref{tab:metadata_not_found}). Importantly, this retrieval rate differs considerably between the two LLMs, which partly explains their difference in accuracy (Table \ref{tab:metadata_not_found}, Fig \ref{fig:rouge_per_field}). In general, Gemini 2.5 Flash has a higher retrieval rate, including for unavailable metadata. We manually examined the 70 instances where Gemini 2.5 Flash `retrieved' unavailable metadata, and found that these were generally sensible and none were totally unfounded (hallucinations). Examples include; access rights labelled as `open access' based on open licenses, authors listed as contact points, (sensible) keywords generated where none were listed, non-specific answers (such as ``Contact owner" as data contact point), or loosely inferred spatial and temporal resolutions (such as `municipalities' or `event-based'). Still, some of these answers were verifiably wrong when the contexts of different metadata were mixed (for example for dates), but most were ambiguous and difficult to score objectively.

\begin{table}[th]
\caption{Fraction of metadata fields that were not identified by the LLMs. For available metadata, this includes structured fields that were not retrieved and unstructured fields that were not recognised (false negatives (FN)) . For unavailable metadata, this includes fields that were reported as unavailable (true negatives (TN)). Post-processed results for both LTER-LIFE and Croissant formats were used.}
\label{tab:metadata_not_found}
\begin{tabular}{lcc}
\toprule
Metadata availability & GPT4 & Gemini 2.5 Flash \\
\midrule
Available and structured metadata not retrieved (FN, $\%$) & 12.5 & 0.0 \\
Available but unstructured metadata not recognised (FN, $\%$) & 21.2 & 6.2 \\
Unavailable metadata reported as unavailable (TN, $\%$) & 56.8 & 34.7 \\
\bottomrule
\end{tabular}
\end{table}

\begin{figure}[th!]
    \centering
    \includegraphics[width=\linewidth]{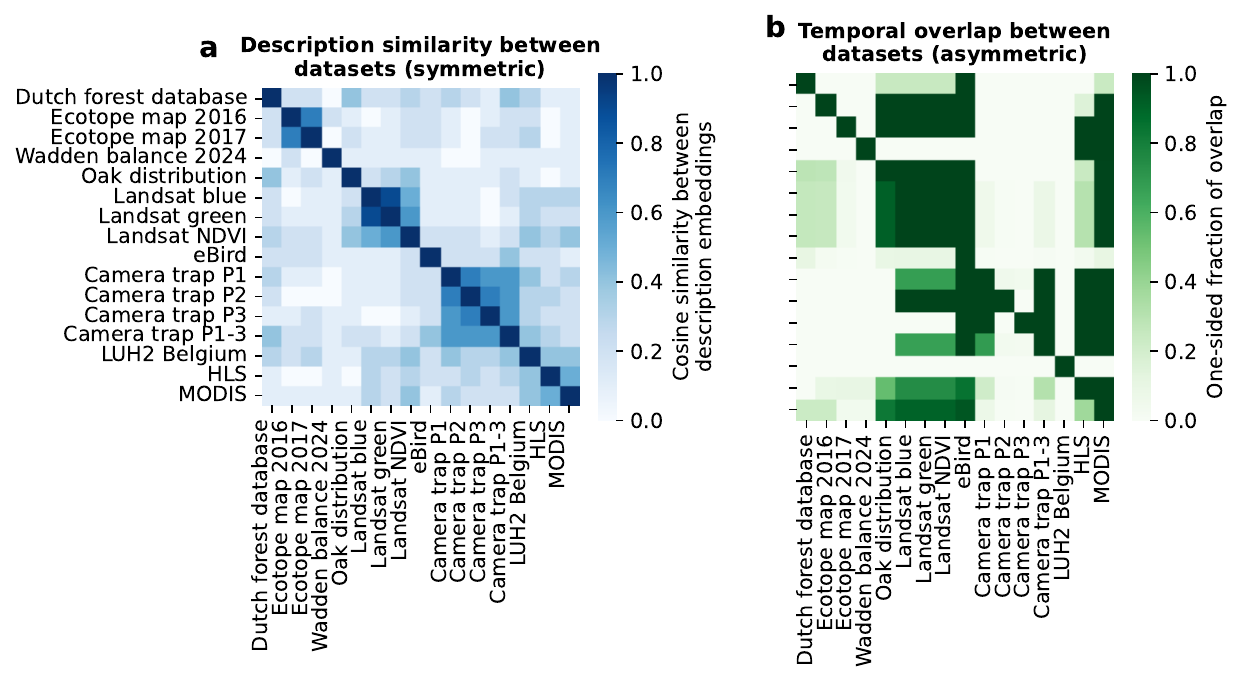}
    \caption{\textbf{a)} Cosine similarity matrix of dataset description embeddings, computed with SBERT sentence embeddings \citep{reimers-2019-sentence-bert}. \textbf{b)} Fraction of temporal overlap, computed one-sided using $\textnormal{fraction}_{ij} = \textnormal{overlap}_{ij} / \textnormal{duration}_i$ for data sets $i$ (y-axis) and $j$ (x-axis).}
    \label{fig:adjacency_desciptions}
\end{figure}

\subsection{Establishing links between datasets}
In this section we aim to demonstrate how the successfully retrieved metadata can be used to create a knowledge base that links datasets from different providers. We trial two methods; first, by directly computing similarity between datasets using LLM embeddings, and second, by using LLMs to convert metadata into a unified format, which can then be used for establishing rule-based knowledge graphs. 
To demonstrate the former, we converted all dataset descriptions into embeddings using the SBERT sentence embeddings \citep{reimers-2019-sentence-bert}, and then calculated pairwise cosine similarity between embeddings (Fig \ref{fig:adjacency_desciptions}a). This similarity matrix successfully identifies clusters of similar datasets -- which we purposefully included in the evaluation dataset (Fig \ref{fig:adjacency_desciptions}a, Table \ref{tab:dataset_overview}). Next, we used an LLM to unify the formatting of extracted \textit{Temporal coverage} metadata (Table \ref{tab:temporal_coverage_conversion}), such that they could be processed using rule-based logic, and as an example, we calculated the temporal overlap between datasets (Fig \ref{fig:adjacency_desciptions}b). Together, these examples illustrate how the extracted metadata can be used to connect datasets for subsequent graph-based RAG and knowledge graph applications. 

\begin{table}[th]
\caption{After the LLM metadata harvester retrieved the \textit{Temporal coverage} of each dataset, another LLM (Gemini 2.5 Flash) call was made to convert these into a consistent format (YYYY-MM-DD-YYYY-MM-DD), where `Present' was instructed to be converted to 2025-06-7.}
\label{tab:temporal_coverage_conversion}
\begin{tabular}{P{3.4cm}P{5.3cm}P{3.2cm}}
\toprule
\textbf{Dataset} & \textbf{LLM-retrieved temporal coverage} & \textbf{LLM-formatted temporal coverage} \\
\midrule
Dutch forest database & 1982-2005 (Measurements), 1983-2000 (Designation) & 1982-01-01-2005-12-31 \\
\midrule
Ecotope map 2016 & 2010-12-08 to 2016-11-01 & 2010-12-08-2016-11-01 \\
Ecotope map 2017 & 2017 & 2017-01-01-2017-12-31 \\
Wadden balance 2024 & 2024 & 2024-01-01-2024-12-31 \\
\midrule
Oak distribution & 2000-01-01 00:00:00 UTC – 2020-12-31 00:00:00 UTC & 2000-01-01-2020-12-31 \\
Landsat blue & 2000-01-01 00:00:00 UTC – 2022-12-31 00:00:00 UTC & 2000-01-01-2022-12-31 \\
Landsat green & 2000-01-01 00:00:00 UTC – 2022-12-31 00:00:00 UTC & 2000-01-01-2022-12-31 \\
Landsat NDVI & 2000-01-01 00:00:00 UTC – 2022-12-31 00:00:00 UTC & 2000-01-01-2022-12-31 \\
\midrule
eBird & January 1, 1800 - December 31, 2023 & 1800-01-01-2023-12-31 \\
Camera trap P1 & August 13th 2021 - August 2023 & 2021-08-13-2023-08-31 \\
Camera trap P2 & August 14, 2021 - September 24, 2021 & 2021-08-14-2021-09-24 \\
Camera trap P3 & March 1, 2023 - March 31, 2023 & 2023-03-01-2023-03-31 \\
\midrule
Camera trap P1-3 & 2021–2023 & 2021-01-01-2023-12-31 \\
LUH2 Belgium & Present to 2050 & 2025-06-07-2050-12-31 \\
\midrule
HLS & 2015-11-28T00:00:00Z–2025-05-31T23:38:19Z & 2015-11-28-2025-05-31 \\
\midrule
MODIS & 2000-02-18 to Present & 2000-02-18-2025-06-07 \\
\bottomrule
\end{tabular}
\end{table}

\section{Discussion}
Searching for datasets remains a major obstacle for researchers \citep{Gregory2020Lost}, and while the introduction of metadata standards has improved data FAIRness, differences in metadata standards and their variable uptake require an adaptive solution to efficiently search for datasets across providers \citep{loffler2021dataset}. Here, we have addressed this challenge by using LLMs to flexibly retrieve and convert both structured and unstructured metadata, which achieve equivalent retrieval accuracy (Fig \ref{fig:rouge_unstructured}). We found that LLM post-processing was key to achieving good performance, followed by choice of LLM (Fig \ref{fig:rouge_per_field}). 
We then demonstrated how LLMs can link datasets using either LLM-based embeddings or by unifying formats for rule-based processing (Fig \ref{fig:adjacency_desciptions}). 

We manually annotated 24 metadata fields for 16 diverse datasets across 7 data providers to validate our approach. Most metadata fields could be evaluated using the deterministic natural language processing metric ROUGE \citep{lin2004rouge}, and additionally we used LLM metrics Faithfulness and Relevancy to evaluate the free-form fields \textit{Description} and \textit{Keywords}. 
When metadata field definitions do not include strict formatting rules, ROUGE can underestimate the accuracy if the correct information is formatted differently than the ground-truth annotations (as exemplified for \textit{Temporal coverage}, Table \ref{tab:temporal_coverage_conversion}).
The most challenging LLM results to validate were instances where metadata was absent (Table \ref{tab:metadata_not_found}), where results were often ambiguous. Ideally, LLM behaviour in these cases should follow user preference, specifying whether to report these metadata as missing or to provide a `best guess' if possible. 

While retrieving metadata solves part of the challenge of searching for datasets across providers, another challenge is linking datasets to create a knowledge base. 
To address this, we prompted an LLM to convert all \textit{Temporal coverage} metadata into a unified format, which then enabled rule-based processing (such as calculating temporal overlap between datasets, Fig \ref{fig:adjacency_desciptions}b). 
Alternatively, we used embedding similarity to quantify connections between datasets for free-form text such as the \textit{Description} field \citep{hayashi2024metadata, sundaram2023making}. In our analysis, this successfully identified clusters of similar datasets that we purposefully included (Fig \ref{fig:adjacency_desciptions}a). This is an important feature, because this can enable researchers to find all datasets similar to one that matches their query, for example via a graph-based RAG system \citep{guo2025lightragsimplefastretrievalaugmented}. 
Here, we constrained the output to contain a single entity per metadata field. While we found that this generally improved performance, a future direction would be to let go of this constraint to enable more flexible links between datasets (\textit{e.g.}, by splitting up data creators into separate entities). 

LLMs are revolutionising science, and promise to have wide-ranging impacts in ecology \citep{rafiq2025generative, reynolds2025potential}. We have developed an LLM application that aids ecological data discovery by flexibly harvesting and organising metadata. We selected a set of 16 diverse datasets to demonstrate and validate our tool, with which a larger, structured knowledge base of datasets can now be constructed. More annotated datasets, including from more languages, would help to validate a broader application of our tool in the future. With different research communities adopting varying metadata standards, we opted for a user-centric design where researchers can specify their desired set of metadata fields and definitions. Our tool hence yields metadata in a unified an standardised format, and a promising future direction is to use this to create formal ontologies or a (graph-based) RAG system to efficiently aid researchers with their data discovery queries \citep{caufield2024structured, hayashi2024metadata}. For example, by specifying metadata fields using the DCAT definitions, extracted metadata could be converted to DCAT ontologies using LLMs or rule-based processing \citep{caufield2024structured, jackson2019robot}. Further, our tool can be integrated into virtual research environments and ecological research infrastructures, for example for automating metadata generation, ensuring compliance to metadata standards or for data discovery. This would reduce the metadata annotation workload for researchers, while enhancing data FAIRness.

\begin{credits}
\subsubsection{\ackname} We acknowledge funding from the Dutch Research Council (NWO) Large-Scale Research Infrastructures (LSRI) programme for the LTER-LIFE (\url{http://www.lter-life.nl}) infrastructure (grant 184.036.014). This version of the contribution has been accepted for publication, after peer review, but is not the Version of Record and does not reflect post-acceptance improvements, or any corrections. The Version of Record is available online at: \url{https://doi.org/10.1007/978-3-032-06136-2_32}.

\subsubsection{Author contributions}
Conceptualization: all authors.
Data curation: PR and TvdP.
Formal analysis: ZL and TvdP.
Funding acquisition: IA and WDK.
Methodology: ZL and TvdP.
Software: ZL and TvdP.
Supervision: IA and WDK.
Validation: ZL and TvdP.
Visualization: TvdP.
Writing – original draft: TvdP.
Writing – review \& editing: all authors.

\subsubsection{\discintname}
The authors have no competing interests to declare that are relevant to the content of this article. 
\end{credits}

\bibliographystyle{splncs04}
\bibliography{main}

\end{document}